# Composite dark matter from 4th generation.

M. Yu. Khlopov*,**[1])

*Center for Cosmoparticle physics "Cosmion", 125047, Moscow, Russia.

**Moscow Engineering Physics Institute (Technical University), Moscow, Russia.

Hypothesis of heavy stable quark of 4th family can provide a nontrivial solution for cosmological dark matter if baryon asymmetry in 4th family has negative sign and the excess of $\bar{U}$ antiquarks with charge (-2/3) is generated in early Universe. Excessive $\bar{U}$ antiquarks form $(\bar{U}\bar{U}\bar{U})$ antibaryons with electric charge -2, which are all captured by $^4He$ and trapped in $[^4He^{++}(\bar{U}\bar{U}\bar{U})^{--}]$ O-helium $OHe$ "atom", as soon as $^4He$ is formed in Big Bang Nucleosynthesis. Interaction of O-helium with nuclei opens new path to creation heavy nuclides in Big Bang nucleosynthesis. Due to large mass of $U$ quark, $OHe$ "atomic" gas decouples from baryonic matter and plays the role of dark matter in large scale structure formation with structures in small scales being suppressed. Owing to nuclear interaction with matter cosmic O-helium from galactic dark matter halo are slowed down in Earth below the thresholds of underground dark matter detectors. However, experimental test of this hypothesis is possible in search for $OHe$ in balloon-borne experiments and for $U$ hadrons in cosmic rays and accelerators. $OHe$ "atoms" might form anomalous isotopes and can cause cold nuclear transformations in matter, offering possible way to exclude (or prove?) their existence.

PACS: *

The problem of existence of new families of quarks and leptons is among the most important in the modern high energy physics. If these quarks and/or leptons are stable, they should be present around us and the reason for their evanescent nature should be found. Recently at least three elementary particle frames for heavy stable charged quarks and leptons were considered: (a) A heavy quark and heavy neutral lepton (neutrino with mass above half the Z-Boson mass) of fourth generation [1]; see also [2], [3]; (b) A Glashow's "Sinister" heavy quark and heavy charged lepton family, bound in "atoms" to be the dominant dark matter [4, 5] and (c) AC-leptons, which are predicted in the extension [7] of standard model, based on the approach of almost-commutative geometry [6], can form evanescent AC-atoms, playing the role of dominant dark matter [7, 8].

The approaches (b) and (c) try to escape the problems of free charged dark matter particles [9] by hiding opposite-charged particles in atom-like bound systems, which interact weakly with baryonic matter. However, in the case of charge symmetry, when primordial abundances of particles and antiparticles are equal, annihilation in early Universe suppresses their concentration. If this primordial abundance still permits these particles and antiparticles to be the dominant dark matter, explosive nature of such dark matter is ruled out by constraints on the products of annihilation in the modern Universe [3, 8]. Even in case of charge asymmetry with primordial particle excess, when there is no annihilation in the modern Universe, binding of positive and negative charge particles is never complete and positively charged heavy species should retain. Recombining with ordinary electrons these heavy positive species give rise to cosmological abundance of anomalous isotopes, exceeding experimental upper limits. To satisfy these upper limits, anomalous isotope abundance in Earth should be reduced, and the mechanisms for such reduction are accompanied by effects of energy release which are strongly constrained, in particular, by the data from large volume detectors.

Here we study the possibility to avoid the problems of composite dark matter models [4, 7], revealed in [3, 5, 8]. We propose a dark matter candidate, which can arise in the model [3], if the baryon asymmetric Universe with normal baryon excess contains also excess of stable antiquarks $\bar{U}$ of 4th generation. In a different framework exotic antibaryon dark matter was recently discussed in [10]. Owing to $\bar{U}$ excess only -2 charge or neutral hadrons are present in the Universe and $^4He$ after it is formed in Big Bang nucleosynthesis completely screens $Q^{--}$ charged hadrons in composite $[^4He^{++}Q^{--}]$ "atoms". These neutral primordial nuclear interacting objects saturate the modern dark matter density and play the role of nontrivial form of strongly interacting dark matter [11, 12]. Active influence of this dark matter on nuclear transformations seem to be incompatible with expected dark matter properties. However, it turns out that the considered scenario is not easily ruled out and deserves attention.

---
[1])e-mail:Maxim.Khlopov@roma1.infn.it

For the quark with electric charge $q = 2/3$ the experimental lower limit is $m_U > 220 \, \text{GeV}$ [13] and we assume that its mass is equal to $m_U = 350 S_5 \, \text{GeV}$. This quark can form lightest $(Uud)$ baryon, $(U\bar{u})$ and corresponding antiparticles are formed by $\bar{U}$ with light antiquarks $\bar{u}$. Owing to large chromo-Coulomb binding energy ($\propto \alpha_c^2 \cdot m_U$, where $\alpha_c$ is the QCD constant) stable double and triple $U$ bound states $(UUq)$, $(UUU)$ and their antiparticles $(\bar{U}\bar{U}\bar{u})$, $(\bar{U}\bar{U}\bar{U})$ can exist [3, 4, 5]. Formation of these double and triple states in particle interactions at accelerators and in cosmic rays is strongly suppressed, but they can form in early Universe and strongly influence cosmological evolution of 4th generation hadrons. As we show, anti-U-triple state called anutium or $\Delta_{3\bar{U}}^{--}$ is of special interest. This stable anti-delta-isobar, composed of $\bar{U}$ antiquarks and bound chromo-Coulomb force has the size $r_\Delta \sim 1/(\alpha_{QCD} \cdot m_U)$, which is much less than normal hadronic size $r_h \sim 1/m_\pi$.

The model [3] admits that in the early Universe a antibaryon asymmetry for 4th generation quarks can be generated so that a $\bar{U}$ excess corresponds to the modern dark matter density. Following [4, 5, 8], it is convenient to relate baryon $\Omega_b = 0.044$ and $\bar{U}$-antibaryon densities $\Omega_{\bar{U}} = \Omega_{CDM} = 0.224$ with the entropy density $s$ and to introduce $r_b = n_b/s$ and $r_{\bar{U}} = n_{\bar{U}}/s$. One obtains $r_b \sim 8 \cdot 10^{-11}$ and $r_{\bar{U}}$, corresponding to $\bar{U}$ excess in the early Universe $\kappa_{\bar{U}} = r_{\bar{U}} - r_U = 10^{-12}(350 \, \text{GeV}/m_U) = 10^{-12}/S_5$, where $S_5 = m_U/350 \, \text{GeV}$.

In the early Universe at temperatures highly above their masses $\bar{U}$ were in thermodynamical equilibrium with relativistic plasma. It means that at $T > m_U$ the excessive $\bar{U}$ were accompanied by $U\bar{U}$ pairs. Their successive evolution after $\bar{U}$ and $U$ freezing out at $T < m_U$ follows the trend studied in details for heavy quarks in [3, 5]. Due to $\bar{U}$ excess frozen out concentration of deficit $U$-quarks is suppressed. It decreases further exponentially first at $T \sim I_U \approx \bar{\alpha}^2 M_U/2 \sim 3 GeV S_5$ (where [3] $\bar{\alpha} = C_F \alpha_c = 4/3 \cdot 0.144 \approx 0.19$ and $M_U = m_U/2$ is the reduced mass), when the frozen out $U$ quarks begin to bind with antiquarks $\bar{U}$ into charmonium-like state $(\bar{U}U)$ and annihilate. On this line $\bar{U}$ excess binds at $T < I_U$ by chromo-Coulomb forces dominantly into $(\bar{U}\bar{U}\bar{U})$ anutium states with mass $m_o = 1.05 \, \text{TeV} S_5$, while remaining free $\bar{U}$ anti-quarks and anti-diquarks $(\bar{U}\bar{U})$ form after QCD phase transition normal size hadrons $(\bar{U}u)$ and $(\bar{U}\bar{U}\bar{u})$. Then at $T = T_{QCD} \approx 150 \, \text{MeV}$ additional suppression of remaining $U$-quark hadrons takes place in their hadronic collisions with $\bar{U}$-hadrons, in which $(\bar{U}U)$ states are formed and $U$-quarks successively annihilate. To the period of Standard Big Bang Nucleosynthesis (SBBN) $\bar{U}$ are dominantly bound in anutium $\Delta_{3\bar{U}}^{--}$) with small fraction ($\sim 10^{-6}$) of neutral $(\bar{U}u)$ and doubly charged $(\bar{U}\bar{U}\bar{u})$ hadron states.

At $T < I_o = Z^2 Z_{He}^2 \alpha^2 m_{He}/2 \approx 1.6 \, \text{MeV}$ the reaction

$$\Delta_{3\bar{U}}^{--} + {}^4He \to \gamma + ({}^4He^{++}\Delta_{3\bar{U}}^{--}) \quad (1)$$

might take place, but it can go only after ${}^4He$ is formed in SBBN at $T < 100 \, \text{keV}$ and is effective only at $T \leq T_{rHe} \sim I_o/\log{(n_\gamma/n_{He})} \approx I_o/27 \approx 60 \, \text{keV}$, when the inverse reaction of photo-destruction cannot prevent it [5, 8]. Since $r_{He} = 0.1 r_b \gg r_\Delta = r_{\bar{U}}/3$, in this reaction all free negatively charged particles are bound with helium [5, 8] and neutral O-helium $({}^4He^{++}\Delta_{3\bar{U}}^{--})$ "atom" is produced with mass $m_{OHe} \approx m_o \approx 1 \, \text{TeV} S_5$. The size of this "atom" is

$$R_o \sim 1/(Z_E Z_{He} \alpha m_{He}) \approx 2 \cdot 10^{-13} \, \text{cm} \quad (2)$$

and it can play the role of dark matter and nontrivial catalyzing role in nuclear transformations.

O-helium looks like an $\alpha$ particle with shielded electric charge. It can closely approach nuclei due to the absence of a Coulomb barrier. On that reason in the presence of O-helium the character of SBBN processes can change drastically.

The size of O-helium is of the order of the size of ${}^4He$ and for a nucleus A with electric charge $Z > 2$ the size of the Bohr orbit for a $(Z\Delta)$ ion is less than the size of nucleus A. This means that while binding with a heavy nucleus $\Delta$ penetrates it and effectively interacts with a part of the nucleus with a size less than the corresponding Bohr orbit. This size corresponds to the size of ${}^4He$, making O-helium the most bound $(Z\Delta)$-atomic state.

The cross section for $\Delta$ interaction with hadrons is suppressed by factor $\sim (p_h/p_\Delta)^2 \sim (r_\Delta/r_h)^2 \approx 10^{-4}/S_5^2$, where $p_h$ and $p_\Delta$ are quark transverse momenta in normal hadrons and in anutium, respectively. Therefore anutium component of $(OHe)$ can hardly be captured and bound with nucleus due to strong interaction.

However, interaction of the ${}^4He$ component of $(OHe)$ with a ${}^A_Z Q$ nucleus can lead to a nuclear transformation due to the reaction

$${}^A_Z Q + (\Delta He) \to {}^{A+4}_{Z+2} Q + \Delta, \quad (3)$$

provided that the masses of the initial and final nuclei satisfy the energy condition

$$M(A, Z) + M(4, 2) - I_o > M(A + 4, Z + 2), \quad (4)$$

where $I_o = 1.6\,\text{MeV}$ is the binding energy of O-helium and $M(4, 2)$ is the mass of the $^4He$ nucleus. The final nucleus is formed in the excited $[\alpha, M(A, Z)]$ state, which can rapidly experience $\alpha$-decay, giving rise to $(OHe)$ regeneration and to effective quasi-elastic process of $(OHe)$-nucleus scattering. It leads to possible suppression of the nuclear transformation (3).

The condition (4) is not valid for stable nuclei participating in reactions of the SBBN. However, unstable tritium $^3H$, produced in SBBN and surviving 12.3 years after it, can react with O-helium, forming $^7Li$ in process $^3H + (^4He\Delta) \to{}^7 Li + \Delta$. Anutium $\Delta^{--}_{3\bar{U}}$, released in this process, is captured by $^4He$ and regenerates O-helium, while $^7Li$ reacts with O-helium, forming $^{11}B$ etc. After $^{39}K$ the chain of transformations starts to create unstable isotopes and gives rise to an extensive tree of transitions along the table of nuclides. This set of processes involves the fraction of baryons of the order of SBBN tritium abundance $(^3H/H \sim 10^{-7})$ and since it does not stop on lithium, but goes further to nuclides, which are observed now with much higher abundance, it can not be excluded by some simple argument. This picture opens new path of chemical evolution of matter on the pre-galactic stage and needs self-consistent consideration within a complete network of nuclear processes.

Note that "atoms" $[^4He^{++}(\bar{U}\bar{U}\bar{u})^{--}]$, which are formed together with O-helium, can catalyze additional recombination of $\bar{U}$-hadrons in anutium in their mutual collisions and in collisions with $(\bar{U}u)$, reducing the fraction of $\bar{U}$-hadrons down to $\sim 10^{-8}$[2].

At $T < T_{od} \approx 1\,\text{keV}$ energy and momentum transfer from baryons to O-helium is not effective $n_b \langle\sigma v\rangle (m_p/m_o)t < 1$. Here

$$\sigma \approx \sigma_o \sim \pi R_o^2 \approx 10^{-25}\,\text{cm}^2. \quad (5)$$

and $v = \sqrt{2T/m_p}$ is baryon thermal velocity. Then O-helium gas decouples from plasma and radiation and plays the role of dark matter, which starts to dominate in the Universe at $T_{RM} = 1\,\text{eV}$.

Development of gravitational instabilities of O-helium gas triggers large scale structure formation, and the composite nature of O-helium makes it more close to warm dark matter.

The total mass of $(OHe)$ within the cosmological horizon in the period of decoupling is independent of $S_5$ and given by

$$M_{od} = \frac{T_{RM}}{T_{od}} m_{Pl} \left(\frac{m_{Pl}}{T_{od}}\right)^2 \approx 2 \cdot 10^{42}\,\text{g} = 10^9 M_\odot. \quad (6)$$

[2] Though binding of these hadrons with nuclei seems unlikely [3], it needs special study and might lead to additional problems.

O-helium is formed only at $T_o = 60\,\text{keV}$ and the total mass of $OHe$ within cosmological horizon in the period of its creation is $M_o = M_{od}(T_o/T_{od})^3 = 10^{37}\,\text{g}$. Though after decoupling Jeans mass in $(OHe)$ gas falls down $M_J \sim 3 \cdot 10^{-14} M_{od}$ one should expect strong suppression of fluctuations on scales $M < M_o$ as well as adiabatic damping of sound waves in RD plasma for scales $M_o < M < M_{od}$. It provides suppression of small scale structure in the considered model.

The cross section of mutual collisions of O-helium "atoms" is given by Eq.(5). $(OHe)$ "atoms" can be considered as collision-less gas in clouds with a number density $n_o$ and size $R$, if $n_o R < 1/\sigma_o$. This condition is valid for O-helium gas in galaxies.

Mutual collisions of O-helium "atoms" determine the evolution timescale for a gravitationally bound system of collision-less $(OHe)$ gas

$$t_{ev} = 1/(n\sigma_o v) \approx 2 \cdot 10^{20} (1\,\text{cm}^{-3}/n)^{7/6}\,\text{s},$$

where the relative velocity $v = \sqrt{GM/R}$ is taken for a cloud of mass $M_o$ and an internal number density $n$. This timescale exceeds substantially the age of the Universe and the internal evolution of O-helium clouds cannot lead to the formation of dense objects.

The first evident consequence of the proposed scenario is the inevitable presence of O-helium in terrestrial matter, which is opaque for $(OHe)$ and stores all its in-falling flux.

If $(OHe)$ capture by nuclei is not effective, its diffusion in matter is determined by elastic collisions, which have transport cross section per nucleon

$$\sigma_{tr} = \pi R_o^2 \frac{m_p}{m_o} \approx 10^{-28}/S_5\,\text{cm}^2. \quad (7)$$

In atmosphere $n_b \sigma_{tr} L_{atm} = 6 \cdot 10^{-2}/S_5$ and the in-falling $(OHe)$ is slowed down in $160 S_5$ m of water (or $40 S_5$ m of rock) and then drifts with velocity $V = \frac{g}{n\sigma v} \approx 80 S_5 A^{1/2}\,\text{cm}/\text{s}$ (where $A \sim 30$ is averaged atomic weight in terrestrial surface matter), sinking down the center of Earth on the timescale $t = R_E/V \approx 1.5 \cdot 10^6 S_5^{-1}\,\text{s}$.

The in-falling O-helium flux from dark matter halo is $I_o = n_o v_h/8\pi$, where number density of $(OHe)$ in vicinity of Solar System is $n_o = 3 \cdot 10^{-4} S_5^{-1}\,\text{cm}^{-3}$ and averaged velocity $v_h \approx 3 \cdot 10^7\,\text{cm}/\text{s}$. During the age of Earth ($t_E \approx 10^{17}\,\text{s}$) about $2 \cdot 10^{38}$ O-helium atoms were captured. If $(OHe)$ dominantly sinks down the Earth, it should be concentrated near the Earth's center within the radius $R_{oc} \sim \sqrt{3T_c/(m_o 4\pi G \rho_c)}$, which is $\leq 3 \cdot 10^7 S_5^{-1/2}\,\text{cm}$ for Earth's central temperature $T_c \sim 10^4\,\text{K}$ and density $\rho_c \sim 4\,\text{g}/\,\text{cm}^3$. Near the Earth's surface O-helium abundance is determined by equilibrium between the in-falling and down-drifting fluxes. It gives $n_o = 2\pi I_o/V = 27 \cdot A^{-1/2}$ cm$^{-3}$, or for $A \sim 30$ about 5 per cm$^{-3}$, being $r_o \sim 5 \cdot 10^{-23}$ relative to number density of terrestrial atoms.

O-helium can be destroyed in reactions (3) [3]. Then free $\Delta^{--}_{3\bar{U}}$ are released and owing to a hybrid Auger effect (capture of $\Delta$ and ejection of ordinary $e$ from the atom with atomic number $A$ and charge of $Z$ of the nucleus) anutium atoms are formed, in which nucleus occupies a highly excited level of the $Z - \Delta$ system, being much deeper than the lowest electronic shell of the considered atom. $\Delta$-atomic transitions to lower-lying states cause radiation in the range intermediate between atomic and nuclear transitions. In course of this falling down to the center of the $Z - \Delta$ system, the nucleus approaches anutium. For $A > 3$ the energy of the lowest state $n$ (given by $E_n = \frac{M\bar{\alpha}^2}{2n^2} = \frac{2Am_p Z^2 \alpha^2}{n^2}$) of the $Z - \Delta$ system (having reduced mass $M \approx Am_p$) with a Bohr orbit, $r_n = \frac{n}{M\bar{\alpha}} = \frac{n}{2AZm_p\alpha}$, exceeding the size of nucleus, $r_A \sim A^{1/3} m_\pi^{-1}$, is less, than the binding energy of $(OHe)$. Therefore regeneration of O-helium in a reaction, inverse to (3), might take place. If regeneration is not effective and $\Delta$ remains bound with heavy nucleus, anomalous isotope of $Z - 2$ element appears. This is the serious problem for the considered model. However, if the general picture of sinking down is valid, it might give no more than $r_o \sim 5 \cdot 10^{-23}$ anomalous isotopes around us, being below experimental upper limits for elements with $Z \geq 2$.

In underground detectors $(OHe)$ "atoms" are slowed down to thermal energies and give rise to energy transfer $\sim 2.5 \cdot 10^{-4}$ eV$A/S_5$ far below the threshold for direct dark matter detection. However, $(OHe)$ destruction can result in observable effects.

O-helium gives rise to less than 0.1 of expected background events in XQC experiment [14], thus avoiding severe constraints on SIMPs obtained in [12] from the results of this experiment.

Atom-like O-helium and tripple-heavy-antiquark anutium can hardly be produced at accelerators, but the proposed in [3] search for $U$ (and $\bar{U}$) hadrons in Run II Tevatron data and in LHC becomes *experimentum crucis* for their basic $\bar{U}$ constituent.

Galactic cosmic rays destroy O-helium. It can lead to appearance of a free anutium component in cosmic rays, which can be as large as $\Delta^{--}_{3\bar{U}}/^4He \sim 10^{-7}$ and accessible to PAMELA and AMS experiments.

The proposed scenario is the minimal for composite dark matter. It assumes only existence of heavy stable $U$-quark and of $\bar{U}$ excess, generated in the early Universe to saturate modern dark matter density. Most of its signatures are determined by nontrivial application of known physics. It might be too simple and too pronounced to be real. In respect to nuclear transformations O-helium looks like the "philosopher's stone", the alchemist's dream. That might be the main reason, why it can not exist. However, its exciting properties make us to remind Voltaire: "Se O-helium n'existai pas, il faudrai l'inventer."

I am grateful to K. M. Belotsky, D. Fargion, M. G. Ryskin, A. A. Starobinsky, C. Stephan and I. I. Tkachev for discussions, to D. Rouable for help and to CRTBT-CNRS and LPSC, Grenoble, France for hospitality.

1. M. Yu. Khlopov, K. I. Shibaev, Gravitation and Cosmology **8**, Suppl., 45 (2002); K. M. Belotsky, M. Yu. Khlopov, K. I. Shibaev, Gravitation and Cosmology **6**, Suppl., 140 (2000); D. Fargion et al, JETP Letters **69**, 434 (1999), astro-ph/9903086; D. Fargion et al, Astropart. Phys. **12**, 307 (2000), astro-ph/9902327; K. M. Belotsky, M. Yu. Khlopov, Gravitation and Cosmology **8**, Suppl., 112 (2002);**7**, 189 (2001).
2. M. Maltoni et al., Phys. Lett. **B476**, 107 (2000); V. A. Ilyin et al., Phys. Lett. **B503**, 126 (2001); V. A. Novikov et al., Phys. Lett. **B529**, 111 (2002); JETP Lett. **76**, 119 (2002).
3. K. M. Belotsky et al., hep-ph/0411271.
4. S. L. Glashow, hep-ph/0504287; A. G. Cohen and S. L. Glashow, in preparation.
5. D. Fargion and M. Khlopov, hep-ph/0507087.
6. A. Connes, *Noncommutative Geometry*, Academic Press, London and San Diego (1994)
7. C. Stephan, hep-th/0509213
8. D. Fargion, M. Yu. Khlopov, C. Stephan, in preparation.
9. S. Dimopoulos et al., Phys. Rev. **D41**, 2388 (1990).
10. D. H. Oaknin and A. Zhitnitsky, Phys. Rev. **D71**, 023519 (2005), hep-ph/0309086; G. R. Farrar and G. Zaharijas, hep-ph/0406281; hep-ph/0510079.
11. C. B. Dover, T. K. Gaisser and G. Steigman, Phys. Rev. Lett. **42**, 1117 (1979); S. Wolfram, Phys. Lett. **B82**, 65 (1979); G. D. Starkman et al., Phys. Rev. **D41**, 3594 (1990); D. Javorsek et al., Phys. Rev. Lett. **87**, 231804 (2001); S. Mitra, Phys. Rev. **D70**, 103517 (2004), astro-ph/0408341.
12. B. D. Wandelt et al., astro-ph/0006344; P. C. McGuire and P. J. Steinhardt, astro-ph/0105567; G. Zaharijas and G. R. Farrar, Phys. Rev. **D72**, 083502 (2005), astro-ph/0406531.
13. D. Acosta et al., (CDF collab.) hep-ex/0211064.
14. D. McCammon et al., Nucl. Instr. Meth. **A370**, 266 (1996); D. McCammon et al., Astrophys. J. **576**, 188 (2002), astro-ph/0205012.

---

[3] Such destruction can be suppressed by immediate $(OHe)$ regeneration due to rapid $\alpha$ decay of the excited final nucleus.